\documentclass{sig-alternate}

\usepackage{times}

\newlength{\off}\newlength{\temp}
\newlength{\laboff} \newcounter{stepno}
\newenvironment{pcode}[2][\arabic{stepno}:]{\setcounter{stepno}{0}
    \paragraph*{#2}\def\linelabel{#1}\begin{list}{}{\leftmargin\off\itemsep0pt
    \topsep0pt\parsep0pt\partopsep0pt\labelsep0pt}}{\end{list}}
\newcommand{\marg}[1][1em]{\end{list}\addtolength{\off}{#1}\begin{list}{}
    {\leftmargin\off\itemsep0pt\topsep0pt\parsep0pt\partopsep0pt\labelsep0pt}}
\newcommand{\step}[1][\linelabel]{\stepcounter{stepno}\settowidth{\temp}
    {{\scriptsize#1}}\item[\hskip-\off\hskip\laboff{\scriptsize#1}\hskip-\laboff
    \hskip-\temp\hskip\off]}

\newtheorem{theorem}{Theorem}[section]
\newtheorem{lemma}[theorem]{Lemma}
\newtheorem{claim}[theorem]{Claim}

\def\N{{\sf I\hspace*{-1pt}N}}
\def\CC{\mathcal{C} }
\def\CS{\mathcal{S} }
\def\CT{\mathcal{T} }
\def\CU{\mathcal{U} }

\conferenceinfo{SPAA'04,} {June 27--30, 2004, Barcelona, Spain.}

\CopyrightYear{2004}

\crdata{1-58113-840-7/04/0006}

\begin{document}

\title{The Effect of Faults on Network Expansion}

\author{
  Amitabha Bagchi
    \titlenote{Dept. of Computer
    Science, University of California, Irvine, CA 92697. {\tt
    \{bagchi,amic\}@ics.uci.edu.}}\\
  \and Ankur Bhargava\titlenote{Dept. of Computer Science, Johns Hopkins
    University, Baltimore, MD 21218. {\tt ankur@cs.jhu.} Supported by NSF Grant CCR-0311321.}
  \and Amitabh Chaudhary{\Large\footnotemark[1]}
  \and David Eppstein\titlenote{Dept. of Computer
    Science, University of California, Irvine, CA 92697. {\tt
  eppstein@ics.uci.edu}. Supported by
  NSF grant CCR-9912338}
  \and Christian Scheideler\titlenote{Dept. of Computer Science, Johns Hopkins
    University, Baltimore, MD 21218. {\tt scheideler@cs.jhu.edu}. Supported by NSF
    grant CCR-0311121 and NSF grant CCR-0311795.}
}

\date{}

\maketitle


\begin{abstract}
In this paper we study the problem of how resilient networks are to
node faults. Specifically, we investigate the question of how many
faults a network can sustain so that it still contains a large
(i.e. linear-sized) connected component that still has approximately
the same expansion as the original fault-free network. For this we
apply a pruning technique which culls away parts of the faulty network
which have poor expansion. This technique can be applied to both
adversarial faults and to random faults. For adversarial faults we
prove that for every network with expansion $\alpha$, a large
connected component with basically the same expansion as the original
network exists for up to a constant times $\alpha \cdot n$
faults. This result is tight in the sense that every graph $G$ of size
$n$ and uniform expansion $\alpha(\cdot)$, i.e. $G$ has an expansion of
$\alpha(n)$ and every subgraph $G'$ of size $m$ of $G$ has an
expansion of $O(\alpha(m))$, can be broken into sublinear components
with $\omega(\alpha(n) \cdot n)$ faults.

For random faults we observe that the situation is significantly
different, because in this case the expansion of a graph only gives a
very weak bound on its resilience to random faults. More specifically,
there are networks of uniform expansion $O(\sqrt{n})$ that are
resilient against a constant fault probability but there are also
networks of uniform expansion $\Omega(1/\log n)$ that are not
resilient again\-st a $O(1/\log n)$ fault probability. Thus, a
different parameter is need\-ed. For this we introduce the {\em span}
of a graph which allows us to determine the maximum fault probability
in a much better way than the expansion can. We use the span to show
the first known results for the effect of random faults on the
expansion of $d$-dimensional meshes.
\end{abstract}

\category{C.2}{Computer Systems Organization}{Computer Communication Networks}
\category{G.2.2}{Mathematics of Computing}{Discrete Mathematics}[Graph Theory]

\terms{Theory, Reliability}

\keywords{faulty networks, expansion, $d$-dimensional mesh, random faults}


\section{Introduction}
\label{sec_intro}

Communication in faulty networks is a classical field in network
theory. In practice, one cannot expect nodes or communication links to
work without complications. Software or hardware faults (or phenomena
outside the control of a network operator such as caterpillars) may
cause nodes or links to go down.  To be able to adapt to faults
without a serious degradation of the service, networks and routing
protocols have to be set up so that they are
fault-tolerant. Fault-tolerant routing has recently attained renewed
interest due to the tremendous rise in popularity of mobile ad-hoc
networks and peer-to-peer networks. In these networks, faults are
actually not an exception but a frequently occurring event: in mobile
ad-hoc networks, users may run out of battery power or may move out of
reach of others, and in peer-to-peer networks, users may leave without
notice.

Central questions in the theoretical area of faulty networks have
been:

\begin{itemize}
\item How many faults can a network sustain so that the size of its
largest connected component is still a constant fraction of the
original size?
\item How many faults can a network sustain so that it can still
emulate its ideal counterpart with constant slowdown?
\end{itemize}

The first question has been heavily studied in the graph theory
community, and the second question has been investigated mostly by the
parallel computing community to find out up to which point a faulty
parallel computer can still emulate an ideal parallel computer with
the same topology with constant slowdown. We refer the reader to
\cite{Sch:02} for a survey of results in these areas.

\subsection{Large connected components in faulty networks}
\label{sec_concomp}

We start with an overview of previous results for random faults and
afterwards consider adversarial faults.

Given a graph $G$ and a probability value $p$, let $G^{(p)}$ be the
random graph obtained from $G$ by keeping each edge of $G$ alive with
probability $p$ (i.e.~$p$ is the {\em survival probability}). Given a
graph $G$, let $\gamma(G) \in [0,1]$ be the fraction of nodes of $G$
contained in a largest connected component.

Let ${\cal G} = \{G_n \mid n \in \N\}$ be any family of graphs with
parameter $n$. Let $p^*$ be the {\em critical probability} for the
existence of a linear-sized connected component. I.e.~for every
constant $\epsilon>0$ it holds:
\begin{enumerate}
\item For every $p>(1+\epsilon)p^*$ there exists a constant $c>0$ with $\lim_{n
\rightarrow \infty} \Pr[\gamma(G^{(p)}_n) > c] = 1$.

\item For all constants $c>0$ and for all $p < (1-\epsilon)p^*$ it holds that
$\lim_{n \rightarrow \infty} \Pr[\gamma(G^{(p)}_n) > c] = 0$.
\end{enumerate}
Of course, it is not obvious whether critical probabilities exist. However, the
results by Erd\H{o}s and R\'enyi \cite{ER60} and its subsequent improvements
(e.g.~\cite{Bo84,LPW94}) imply that for the complete graph on $n$ nodes, $p^* =
1/(n-1)$, and that for a random graph with $d \cdot n/2$ edges, $p^* = 1/d$.
For the 2-dimensional $n \times n$-mesh, Kesten showed that $p^* = 1/2$
\cite{Ke80}. Ajtai, Koml\'os and Szemer\'edi proved that for the hypercube of
dimension $n$, $p^*=1/n$ \cite{AKS82}. For the $n$-dimensional butterfly
network, Karlin, Nelson and Tamaki showed that $0.337 < p^* < 0.436$
\cite{KNT94}. Leighton and Maggs \cite{LM92} showed that there is an indirect
constant-degree network connecting $n$ inputs with $n$ outputs via $\log n$
levels of $n$ nodes each, called multibutterfly, that has the following
property: Up to a constant fault probability it is still possible to find
$O(\log n)$ length paths from a constant fraction of the inputs to a constant
fraction of the outputs. Subsequently Cole, Maggs and Sitaraman \cite{CMS95}
extended this result for the butterfly.

Adversarial fault models have also been investigated. Leighton and
Maggs \cite{LM92} also showed that no matter how an adversary choos\-es
$f$ nodes to fail, there will be a connected component left in the
multibutterfly with at least $n-O(f)$ inputs and at least $n-O(f)$
outputs. (In fact, one can even still route packets between the inputs
and outputs in this component in almost the same amount of time steps
as in the ideal case.) Subsequently Leighton, Maggs and Sitaraman
\cite{LMS98} extended this result for the butterfly.

Upfal \cite{Up94}, following up on work by Dwork et. al.~\cite{DPPU88}
and Alon and Chung~\cite{AC89}, showed that there is also a direct
constant-degree network on $n$ nodes, a so-called expander, that has
the property: no matter how an adversary chooses $f$ nodes to fail,
there will be a connected component left in it with at least $n-O(f)$
nodes. Both results are optimal up to constants. Upfal uses a pruning
technique to achieve his bound which is similar in spirit to the one
we use. Apart from the fact that Upfal gives a polynomial-time
algorithm for pruning while we do not, the important difference worth
noting is that Upfal's pruning does not guarantee a large component of
good expansion. In fact, to the best of our knowledge there is no
known constant approximation algorithm to determine the expansion of a
graph of unknown topology.

\subsection{Simulation of fault-free networks by faulty networks}

Next we look at the problem of simulating fault-free networks by
faulty networks. Consider the situation that there can be up to $f$
worst-case node faults in the system at any time. One way to check
whether the largest remaining component still allows efficient
communication is to check whether it is possible to embed into the
largest connected component of a faulty network a fault-free network
of the same size and kind. An {\em embedding} of a graph $G$ into a
graph $H$ maps the nodes of $G$ to non-faulty nodes of $H$ and the
edges of $G$ to non-faulty paths in $H$. An embedding is called {\em
static} if the mapping of the nodes and edges is fixed. Both static
and dynamic embeddings have been used. A good embedding is one with
minimum load, congestion, and dilation, where the {\em load} of an
embedding is the maximum number of nodes of $G$ that are mapped to any
single node of $H$, the {\em congestion} of an embedding is the
maximum number of paths that pass through any edge $e$ of $H$, and the
{\em dilation} of an embedding is the length of the longest path. The
load, congestion, and dilation of the embedding determine the time
required to emulate each step of $G$ on $H$. In fact, Leighton, Maggs,
and Rao have shown \cite{LMR94} that if there is an embedding of $G$
into $H$ with load $\ell$, congestion $c$, and dilation $d$, then $H$
can emulate any communication step (and also computation step) on $G$
with slowdown $O(\ell+c+d)$.

When demanding a constant slowdown, only a few results are known so
far. In the case of worst-case faults, it was shown by Leighton, Maggs
and Sitaraman (using dynamic embedding strategies) that an $n$-input
butterfly with $n^{1-\epsilon}$ worst-case faults (for any constant
$\epsilon$) can still emulate a fault-free butterfly of the same size
with only constant slowdown \cite{LMS98}.  Furthermore, Cole, Maggs
and Sitaraman showed that an $n \times n$ mesh can sustain up to
$n^{1-\epsilon}$ worst-case faults and still emulate a fault-free mesh
of the same size with (amortized) constant slowdown \cite{CMS97}. It
seems that also the $n$-node hypercube can even achieve a constant
slowdown for $n^{1-\epsilon}$ worst-case faults, but so far only
partial answers have been obtained \cite{LMS98}.

Random faults have also been studied. For example, H{\aa}stad,
Leighton and Newman \cite{HLN87} showed that if each edge of the
hypercube fails independently with any constant probability $p<1$,
then the functioning parts of the hypercube can be reconfigured to
simulate the original hypercube with constant slowdown. Leighton,
Maggs and Sitaraman \cite{LMS98} showed that a butterfly network whose
nodes fail with some constant probability $p$ can still emulate a
fault-free butterfly of the same size with slowdown $2^{O(\log^* n)}$.
Interestingly, in the conference version of~\cite{CMS97}, Cole, Maggs
and Sitaraman claim that an $n \times n$ mesh in which each node is
faulty independently with a constant fault probability is able to
emulate a fault-free mesh with a constant slowdown~\cite{CMS93}. The
proof of this claim, which is stronger than the theorem we prove about
the $n \times n$ mesh in this paper, is omitted in~\cite{CMS93} and
has not appeared elsewhere to the best of our knowledge.

For a list of further references concerning embeddings of fault-free
into faulty networks see the paper by Leighton, Maggs and Sitaraman
\cite{LMS98}.

\subsection{Our approach}

The two common approaches -- connectivity and emulation of fault-free
by faulty networks -- are too extreme for many practical
applications. Knowing how long a network is still connected may not be
very useful, because in extreme cases (just a single line connects one
half to the other) the speed of communication may be reduced to a
crawl, making it useless for applications that need a fast interaction
or a large bandwidth such as interactive gaming or video
conferences. On the other hand, emulating a fault free network
on a faulty network is like using a giant hammer to crack
a lesser nut, so to speak. Emulation may not be needed when
all we want is reduced congestion or good expansion.

Applications in ad-hoc networks or peer-to-peer systems usually do not
care about how a network is connected, concerning themselves instead
with whether it still provides sufficient bandwidth and ensures
sufficiently small delays. In this scenario a more relevant question
is:

\begin{quote}
\em How many faults can a network sustain so that it still contains a
network of at least a constant fraction of its original size that
still has approximately the same expansion?
\end{quote}

Knowing an answer to this question would have many useful consequences
for distributed data management, routing, and distributed
computing. Research on load balancing has shown that if the expansion
basically stays the same, the ability of a network to balance
single-commodity or multi-commodity load basically stays the same, and
this ability can be exploited through simple local
algorithms~\cite{GLMMPRRTZ99,AKK02}.  Also, the ability of a network
to route information is preserved because it is closely related to its
expansion \cite{Sch98}. Furthermore, as long as the original network
still has a large connected component of almost the same expansion,
one can still achieve almost everywhere agreement which is an
important prerequisite for fundamental primitives such as atomic
broadcast, Byzantine agreement, and clock synchronization
\cite{DPPU88,Up94,BR96}.

Many different fault models have been studied in the literature:
faults may be permanent or transient, nodes and/or edges may break
down, and faults may happen at random or may be caused by an adversary
or attacker. The former faults are called {\em random faults}, and the
latter faults are called {\em adversarial faults}. We will concentrate
on situations in which there are static node faults, i.e. nodes either
break down randomly or due to some adversary. For adversarial faults,
we will consider the node expansion of a graph, and for random faults
we will use the edge expansion of a graph.

Given a graph $G=(V,E)$ and a subset $U \subseteq V$, the {\em (node)
expansion} of $U$ is defined as
\[
  \alpha(U) = \frac{|\Gamma(U)|}{|U|}
\]
where $\Gamma(U)$ is the set of nodes in $V \setminus U$ that have an edge from
$U$ and $|S|$ denotes the size of set $S$. The {\em (node) expansion} of $G$ is
defined as $\alpha = \min_{U, |U| \le |V|/2} \alpha(U)$.

Similarly, the edge expansion of $G$ is defined as:
\[
  \alpha_e = \min_{U \subseteq V} \left\{ \frac{|(U,V \setminus U)|}{\min \{
    |U|,|V \setminus U|\}} \right\}
\]
where $(U,V \setminus U)$ denotes the set of edges with one endpoint in $U$ and
the other in $V \setminus U$.

\subsection{Our main results}

\subsubsection*{Adversarial faults}

We give general upper and lower bounds for the number of node faults a graph
can sustain so that it still has a large component with basically the same
expansion, where the bounds are tight up to a constant factor. More
specifically, we show that the number of adversarial node faults a graph with
node expansion $\alpha$ and $n$ nodes can sustain, with only a constant factor
decrease in its expansion, is a constant times $\alpha \cdot n$. For graphs $G$
of size $n$ and uniform expansion $\alpha(\cdot)$, i.e. $G$ has an expansion of
$\alpha(n)$ and every subgraph $G'$ of size $m$ of $G$ has an expansion of
$O(\alpha(m))$, this result is best possible up to constant factors.

\subsubsection*{Random faults}

We also study random faults. Our main contribution here is to suggest a new
parameter for their study, which may be of independent interest.

Consider a graph $G=(V,E)$. Let $U \subseteq V$ be any subset of nodes. $U$ is
defined to be {\em compact} if and only if $U$ and $V \setminus U$ are
connected in $G$. Let $\cal U$ be the set of all compact sets of $G$. Let
$P(U)$ be the smallest tree in $G$ which connects every node in $\Gamma(U)$
(i.e. it essentially spans the boundary of $U$). Note that the set of nodes in
$P(U)$ need not be from $U$ alone or from $V\setminus U$ alone. Then the {\it
span} of a graph is defined as:
\begin{equation}
\sigma = \max_{U \in{\cal U}} \left\{ { |P(U)| \over |\Gamma(U)| } \right\}
\end{equation}
The span helps us characterize the resilience of the expansion to
random faults. We show that a graph with maximum degree $\delta$ and
span $\sigma$ can tolerate a fault probability up to a constant times
${1 \over \delta^\sigma}$ and still retain an expansion within a
factor of $\delta$ of its original expansion.

We also show that the $d$-dimensional meshes have constant span. The
proof of this theorem is of independent value as it establishes an
interesting property of the $d$-dimensional mesh: The boundary of any
set of connected vertices in the $d$-dimensional mesh, whose
complement is also connected, can be spanned by a tree of size at
most twice the size of the boundary.

\subsection{Outline of the paper}

The rest of the paper is organized as follows: In
Section~\ref{sec_adversary} we consider adversarial faults, and in
Section~\ref{sec_random} we consider random faults. The paper ends in
Section~\ref{sec_conclusion} with a discussion of how our results are
related to previous research and some open problems.


\section{Adversarial faults} \label{sec_adversary}

In this section we prove the existence of a large connected component with good
expansion in a graph with faulty nodes.  We assume that a malicious adversary
decides which nodes are faulty.  More formally, we are given a network $G =
(V,E)$ with $n$ nodes and vertex expansion $\alpha$. An adversary gives us a
faulty version of this network, called $G_f$, with $f$ faulty nodes removed. We
will show that there exists a subnetwork of $G_f$ called $H$ which has
$\Theta(n)$ nodes and has an expansion of $\Theta(\alpha)$ provided that the
adversary is given no more than $O(\alpha \cdot n)$ faults.

We cannot argue that the expansion of $G_f$ is no more than a constant factor
less than $\alpha$ for the simple reason that the adversary can create
bottlenecks in the network. However, we describe a way to find a large
connected component of $G_f$ with the required properties using an algorithm called
{\em Prune} described in Figure~\ref{fig:prune}. Note that the running time of
{\em Prune} is not necessarily polynomial, nor are we claiming it is. {\em
Prune} simply helps us prove an existential result.

Before we get to the algorithm we need to introduce some notation. We define
$\Gamma(S)$ to be the set of nodes in the neighbourhood of a subnetwork $S$.
The algorithm generates a sequence of graphs $G_0$ to $G_m$.  We now present
the algorithm and state the main theorem of this subsection.

\begin{figure}[ht]
\begin{pcode}{Algorithm {\em Prune($\epsilon$)}}
\step $G_0 \leftarrow G_f$; $\,i \leftarrow 0$
\step {\bf while } $\exists S_i \subseteq G_i$ such that $|\Gamma(S_i)| \leq
    \alpha \cdot \epsilon \cdot |S_i|$ and $|S_i| \leq |G_i|/2$
\marg
    \step $G_{i+1} \leftarrow G_i \setminus S_i $
    \step $i \leftarrow i + 1$
\marg[-1em]
\step {\bf end while}
\step $H \leftarrow G_i$; $\,m \leftarrow i$
\end{pcode}
\caption{The pruning algorithm}
\label{fig:prune}
\end{figure}

\begin{theorem}
\label{thm:adversarial_expansion}
Given a network $G$ with $n$ nodes, node expansion $\alpha$ and $f$ faulty
nodes chosen by an adversary, for any constant $k$ such that $k \geq 2$ and
$\frac{k \cdot f}{\alpha} \leq \frac{n}{4}$, {\em Prune($1 - \frac{1}{k}$)}
returns a subnetwork $H$ of at least size $n - \frac{f \cdot k}{\alpha}$ with
expansion $(1 - \frac{1}{k}) \cdot \alpha$.
\end{theorem}
\begin{proof}
Denote $G_f \setminus H$ as $\CS$. $\CS$ is thus the union of all the regions
culled by {\em Prune}.  To prove the result we will first show that the size of
$\CS$ is bounded by $\frac{k \cdot f}{ \alpha}$. To show this we will use the
fact that the number of faults required to cull a region is proportional to
the size of the region. To demonstrate that we need the following lemma.

\begin{lemma}
\label{lem:puttogether}
\[
\left|\Gamma(\bigcup_{0 \leq i \leq j} S_i)\right| \leq
\sum_{0\leq i\leq j} |\Gamma (S_i)|\leq \alpha \cdot (1 -
\frac{1}{k}) \cdot \left|\bigcup_{0 \leq i \leq j}S_i\right|.
\]
\end{lemma}
\begin{proof}
Consider the first inequality. Obviously, any node that lies in the
neighborhood of $\bigcup_i S_i$ must lie in the neighborhood of some $S_i$.
Therefore $\Gamma(\bigcup_i S_i) \subseteq \bigcup_i \Gamma(S_i)$. Hence the
first inequality. Each set $S_i$ that is culled by {\em Prune($1-{1\over k}$)}
has the property that $|\Gamma(S_i)| \leq \alpha \cdot (1-{1\over k}) \cdot
|S_i|$. Since the sets $S_i$ are disjoint, $\sum_i |S_i| = |\bigcup_i S_i |$.
Hence the second inequality.
\end{proof}

We will show that $\CS \leq \frac{k\cdot f}{\alpha}$ by contradiction. Let, if
possible, $\CS > \frac{k\cdot f}{\alpha}$. Since at every iteration of the
algorithm we pick an $S_i$ which is the smaller side of the cut we have found,
each $S_i$ is at most $n/2$ in size.  Now, since $\frac{k\cdot f}{\alpha} \leq
\frac{n}{4}$, there is a $j$ such that either $ \frac{k\cdot f}{\alpha} < \left
|\bigcup_{0 \leq i \leq j}S_i \right | \leq n/2$ or $S_j$ such that
$\frac{k\cdot f}{\alpha} < |S_j| \leq n/2$.  So we can always choose an $\CS'
\subseteq \CS$ such that $\frac{k\cdot f}{\alpha} < |\CS'| \leq n/2$.  In
either case, from Lemma \ref{lem:puttogether}, we have:
\[
\Gamma(\CS') \leq \alpha \cdot (1 - \frac{1}{k}) \cdot |\CS'|.
\]
We know that in $G$, $\,|\Gamma(\CS')|$ is at least $\alpha \cdot |\CS'|$.
Hence, the number of faulty nodes in $\CS'$'s neighborhood must be at least
$\alpha(1 - (1 - \frac{1}{k})) \cdot |\CS'|$ i.e. greater than $\alpha \cdot
\frac{1}{k} \cdot \frac{k\cdot f}{\alpha}$ i.e. greater than $f$.  Since the
total number of faults allowed to the adversary is at most this number, we have
a contradiction.  Hence, $H$ is at least $n - \frac{k \cdot f}{\alpha}$ in size
and has expansion at least $(1 - \frac{1}{k}) \cdot \alpha$. \qed
\end{proof}

The result given in Theorem~\ref{thm:adversarial_expansion} is the best
possible up to constant factors. To prove this we will first show that for
every $\alpha>0$ smaller than some constant there is an infinite family of
graphs which disintegrate into sublinear components on removing some $c \cdot
\alpha \cdot n$ vertices where $n$ is the number of nodes in the given graph
and $c$ is some constant. Then we show that
Theorem~\ref{thm:adversarial_expansion} is also the best possible up to
constant factors for arbitrary graphs of uniform expansion.

\begin{theorem}
\label{thm:adv_lowerbound} There exists a constant $\beta$ such that, given any
$\alpha < \beta$, there is an infinite family of graphs with expansion $\alpha$
for which there is an adversarial selection of $c \cdot \alpha \cdot n$ faulty
nodes causing the graph to break into sublinear components, where $n$ is the
number of nodes in the graph and $c$ is an appropriately chosen constant.
\end{theorem}

\begin{proof}
To construct this family of graphs let us consider $G(n)$ to be an infinite
family of expander graphs with constant expansion $\beta$ and constant degree
$\delta$.

For each $G \in G(n)$, construct a graph, $H$, which is a copy of $G$ with each
edge replaced by a chain of $k$ nodes, where $k$ is even. Then $H$ has ${\delta
\cdot n \cdot k \over 2} + n = O(k \cdot n)$ nodes.

\begin{claim}
\label{clm:h_expansion} Graph $H$ has expansion $\Theta({1 \over k})$.
\end{claim}

\begin{proof}
Take any subset $U$ of nodes in $H$ representing original nodes in $G$ and let
$U'$ be the set resulting from $U$ by adding the $k/2$ nearest nodes of each
chain a node in $U$ is connected to. Then $|U'| = (\frac{\delta \cdot k}{2} + 1) \cdot |U|$ but $|\Gamma(U')| = |\Gamma(U)| \le \delta \cdot |U|$. Hence,
\[
  \alpha(U') = \frac{|\Gamma(U')|}{|U'|} \le \frac{2}{k} \cdot |U'|
\]
completing the proof of the claim.
\end{proof}

Now, from each chain of $k$ nodes we remove the central node. Each component
remaining has $\delta \cdot {k \over 2}$ nodes left, i.e. a sublinear number,
and the total number of nodes removed is ${\delta \over 2} \cdot n$, which is
${1 \over k}$ times the number of nodes in the graph. \qed
\end{proof}

Recall that a graph $G$ of size $n$ is of uniform expansion
$\alpha(\cdot)$ if the expansion of $G$ is $\alpha(n)$ and every
subgraph $G'$ of size $m$ of $G$ has an expansion of
$O(\alpha(m))$. This is the case for all well-known classes of
graphs. Consider, for example, the $m \times m$-mesh with $n=m^2$
nodes and let $\alpha(m) = \sqrt{m}$. Its expansion approximately
$\sqrt{n}$, and every subgraph of that mesh of size $m$ has an
expansion of $O(\sqrt{m})$. Hence, it has a uniform expansion.

\begin{theorem}
For every connected graph of size $n$ and uniform expansion $\alpha(x)$
there
is an adversarial selection of $\omega(\alpha(n) \cdot n)$ faulty nodes
that
causes the graph to break into sublinear components.
\end{theorem}
\begin{proof}
Let $G=(V,E)$ be any graph of uniform expansion $\alpha(x)$ that
consists of
$n$ nodes. Then there must be a set $U_1 \subseteq V$, $|U_1| \le n/2$,
so that
$|\Gamma(U_1)| \le \alpha(n) \cdot |U_1|$. Removing $\Gamma(U_1)$ leaves
$G$
with a set ${\cal V}_1 = \{V',V''\}$ of two node sets, $V'=U_1$ and $V''
= V
\setminus (U_1 \cup \Gamma(U_1))$. Let $V_1$ be a set in ${\cal V}_1$ of
maximum size. It follows from the uniformity of $G$ that there must be a
set
$U_2 \subseteq V_1$, $|U_2| \le |V_1|/2$, so that $|\Gamma(U_2)|$ w.r.t.
$G(V_1)$ is $O(\alpha (|V_1|)) \cdot |U_2|$. Removing $U_2$ results in a
new
set ${\cal V}_2$ of sets of nodes in which $V_1$ is replaced by $U_2$
and $V_1
\setminus (U_2 \cup \Gamma(U_2))$. We continue to take a node set $V_i$
of
largest size out of ${\cal V}_i$ and remove nodes at the minimum
expansion part
in $G(V_i)$ until there is no subset in ${\cal V}_i$ left of size at
least
$\epsilon n$.

Our goal is to show that this process only removes $O(\frac{\log
(1/\epsilon)}{\epsilon} \cdot \alpha(n) \cdot n)$ nodes from $G$. If
this is
true, the theorem would follow immediately. We prove the bound with a
charging
strategy: Each time a set $V_i$ is selected from ${\cal V}_i$, we charge
all
nodes in $\Gamma(U_{i+1})$ taken away from $V_i$ to the nodes in
$U_{i+1}$.
Since
\[
  |\Gamma(U_{i+1})| = O(\alpha(\epsilon n)) \cdot |U_{i+1}|
  = O \left( \frac{\alpha(n)}{\epsilon} \cdot |U_{i+1}| \right)
\]
for any $\alpha(x) \ge 1/x$, this means that every node in $U_{i+1}$ is
charged
with a value of $O(\epsilon^{-1} \cdot \alpha(n))$. Every node can be
charged
at most $\log (1/\epsilon)$ times because each time a node is charged,
it ends
up in a node set $U_{i+1}$ that is at most half as large as $V_i$, and
we stop
splitting a node set once it is of size less than $\epsilon n$. Hence,
at the
end, every node in $V$ is charged with a value of $O(\frac{\log
(1/\epsilon)}{\epsilon} \cdot \alpha(n))$. Summing up over all nodes,
the total
charge is
\[
  O \left( \frac{\log(1/\epsilon)}{\epsilon} \cdot \alpha(n) \cdot n
\right) \;
  ,
\]
which represents the number of nodes that have been removed from the
graph.
\end{proof}



\section{Random faults} \label{sec_random}

We now direct our attention to the case of random faults. We assume
that each node in the network can independently become faulty with a
given probability $p$.

\subsection{Random faults aren't (always) easier to handle}

Intuitively it appears that in general this situation might be easier
to handle since there is no malicious adversarial intent behind the
distribution of node failures. But, in general this does not seem to
be true. We begin this section by showing that there are families of
graphs for which a fault probability of $\Theta(\alpha)$ causes the
graph to disintegrate into sublinear fragments, where $\alpha$ is the
node expansion of the graph. In other words, in these graphs
$\Theta(\alpha n)$ random node failures can be catastrophic: they
don't even allow us to find a linear sized connected component, hence
making it impossible to find a linear sized connected component with
good expansion.

To construct this family of graphs we begin with an infinite family of
constant degree expander graphs with a constant node expansion $\beta$ and
maximum degree $\delta$. We denote this family as $G(n)$.

\begin{theorem}
\label{thm:rand_lb}
Given any $\alpha < \beta$, there exists an infinite family of graphs
with node expansion $\alpha$ for which a fault probability of $\frac{3\log
\delta}{\beta} \cdot \alpha$ causes the graph to disintegrate.

\end{theorem}

\begin{proof}
We use the family of graphs constructed in the proof of
Theorem\-~\ref{thm:adv_lowerbound}, i.e. let $G(n)$ be an infinite family of
constant degree expander graphs with constant expansion $\beta$ and degree
$\delta$. Construct a graph, $H$, which is a copy of $G$ with each edge
replaced by a chain of $k$ nodes. Graph $H$ has $O(k \cdot n)$ nodes. From
Claim~\ref{clm:h_expansion} we know that $H$ has expansion
$\Theta(\frac{1}{k})$. Excercise 5.7 of~\cite{MR95} gives us the
following important property of $H$:

\begin{claim}
\label{clm:comp_count}
The number of connected subgraphs of $H$ with $r$ vertices from $G$ in
them is at most $n \cdot \delta^{2r}$.
\end{claim}

\begin{proof}
Any connected
subgraph of size $r$ can be spanned by a tree with $r - 1$ edges. This
tree can be traversed by an Eulerian tour in which each edge is used
at most twice. Hence the subgraph is represented by a walk along the
graph of length at most $2r$ vertices from $G$. Since the root can be
one of $n$ vertices, the result follows.
\end{proof}

Let the failure probability of the nodes in $H$ be $p = { 4 \ln \delta \over k
}$. Consider any subgraph of $H$ with $r = \ln n$ vertices from $G$. The total
number of nodes in this subgraph is at most $\delta \cdot k \cdot r$ and at
least $k \cdot r$.  Hence, this particular subgraph survives in $H$ with
probability at most $(1-p) ^ {k \cdot r} \leq e^{-k \cdot r \cdot p}$.  By
Claim~\ref{clm:comp_count} there are no more that $n \cdot \delta^{2r}$ such
components in $H$. Hence, the probability that such a subgraph survives is at
most $n \cdot \delta^{2r} \cdot e^{-k \cdot r \cdot p} = n^{1- 2 \ln \delta}
\leq {1 \over n}$.  Since with high probability there can be no connected
subgraph with size $\Theta(\delta \cdot k \ln n)$ in $H$ which has $k\cdot n$
vertices and $\delta$ is a constant, we conclude that $H$ breaks down into
sublinear components with high probability.

In the above construction, set $k=\lceil {\beta \over \alpha} \rceil$ for a
given $\alpha < \beta$ and the theorem follows. \qed
\end{proof}

However it isn't as if the expansion of the graph is a critical point
for all graphs. There are several important classes of graphs which
can sustain a much higher fault probability and still yield a linear
sized connected component with good expansion.

\subsection{Extracting a subnetwork of size $\Theta(n)$ and edge expansion
    $\Theta(\alpha_e)$}

We are given a network $G=(V,E)$ with $n$ nodes, edge expansion $\alpha_e$ and
graph span $\sigma$. Let us call the faulty version of this network $G_f$.  We
want to find a network $H \subseteq G_f$ of size $\Theta(n)$ with edge
expansion $\Theta(\alpha_e)$. Let $\cal U$ be the set of all compact sets of
$G$. Note that a set is compact if both it and its complement are connected. We
will use the notion of {\it edge expansion} in this section.

\begin{lemma}
\label{lem:compact} If $S \subset G$ is connected and $|S|<n/2$ then there
exists a compact set $K_G(S)$ in $G$ whose edge expansion is no more than $S$'s
edge expansion.
\end{lemma}

\begin{proof}
If $S \in \CU$ then $K_G(S)$ is simply $S$. If $S \notin \CU$, $G \setminus S$
is not connected. Let $\CC(S)$ be the set of maximal connected subgraphs of $G
\setminus S$. Let $\Gamma_e(\cdot)$ be the set of edges leaving a set. It is clear
that $\CC(S) \subset \CU$ (if not then they are not maximal). We consider two
cases.

\noindent{\em Case 1:} There is a $C\in \CC(S)$ with $|C| \geq n/2$.\\
Then $G \setminus C \in \CU$, $S\subseteq G \setminus C$, $|G \setminus C| <
n/2$, and $\Gamma_e (G \setminus C) \subseteq \Gamma_e (S)$. Hence, $G\setminus
C$ has an edge expansion less than $S$'s edge expansion. So, $K_G(S)=G
\setminus C$.

\noindent{\em Case 2:} For all $C\in \CC(S)$, $|C| < n/2$.\\
If any of the connected components in $\CC(S)$ has a an edge expansion less
than $S$'s then let that component be $K_G(S)$. If not, then all components
$C_i\in \CC(S)$ have an edge expansion strictly larger than $S$'s, i.e. for all
$i$, ${ \Gamma_e (C_i) \over |C_i| } > {\Gamma_e (S) \over |S|}$. But,
$\Gamma_e ( \cup_i C_i) = \Gamma_e (S)$.  Hence, $|S| > |G \setminus S|$, which
is a contradiction. Therefore, one of the $C_i$'s must have an edge expansion
less than or equal to $S$'s edge expansion.
\end{proof}

\begin{figure}[ht]
\begin{pcode}{Algorithm {\em Prune2($\epsilon$)}}
\step $G_0 \leftarrow G_f$; $\,i \leftarrow 0$ \step {\bf while} $\exists
(S_i,G_i\setminus S_i)$ in $G_i$ s.t.
    $|(S_i,G_i \setminus S_i)| \leq \alpha_e \cdot \epsilon \cdot |S_i|$ and
    $|S_i| \leq |G_i|/2$ and $S_i$ is connected
\marg
    \step $K_i \leftarrow K_{G_i}(S_i)$
    \step $G_{i+1} \leftarrow G_i \setminus  K_i$
    \step $i \leftarrow i + 1$
\marg[-1em]
\step {\bf end while}
\step $H \leftarrow G_i$
\end{pcode}
\caption{The pruning algorithm}
\label{fig:prune2}
\end{figure}

We use notation from algorithm {\em Prune2} in the proof and statement of
theorem \ref{thm:randfaults}.

\begin{theorem}
\label{thm:randfaults}
{\em Prune2($\epsilon$)} returns a subnetwork $H$ of size $|H| \geq n/2$ with
edge expansion $\epsilon \cdot \alpha_e$ with high probability, provided that
edge expansion, $\alpha_e \geq { 6 \delta^2 \cdot \log^3_\delta n \over n}$,
fault probability, $p \leq { 1 \over 2 e \cdot \delta^{4\sigma}}$ and
degradation in expansion, $\epsilon \leq {1 \over 2 \delta}$.
\end{theorem}

\begin{proof}

Let $\CT=G_f \setminus H$. Hence $\CT$ is the union of all the culled regions.
To prove the result we will show that with high probability the size of $\CT$
is not more than $n/2$.  Let $\{T_1, T_2, \ldots, T_l\}$ be maximal connected
components of $\CT$.

\begin{claim}
\label{claim:compact}
$\forall T_i \in \CT$, $T_i$ is compact in $G_f$.
\end{claim}

\begin{proof}
Suppose $T_i$ is not compact in $G_f$. Select the largest $j$ such that $T_i$
is not compact in $G_j$ and $T_i \subseteq G_j$. (i.e. no part of $T_i$ has
been culled yet, which means that $G_{j+1}$ is well-defined.) Let us consider
two cases:

\noindent{\em Case 1:} $T_i \subseteq G_{j+1}$\\
This means that $T_i$ must be compact in $G_{j+1}$ else $j$ could have been one
higher. So, we have $3$ components in $G_j$, namely: $K_j$, $T_i$ and $G_{j+1}
\setminus T_i$. Since $T_i$ is noncompact in $G_j$, the neighborhood of $K_j$
in $G_j$ is wholly in  $T_i$.  Since $K_j$ is disjoint with $T_i$, $T_i$ is not
maximal. Contradiction.

\noindent{\em Case 2:} $T_i \not \subseteq G_{j+1}$\\
This means that $T_i$ and $K_j$ are not disjoint. Since $K_j$ is a culled set
it must be wholly inside $T_i$, else $T_i$ is not maximal. $T_i$ is not compact
in $G_j$, so $T_i\setminus K_j$ is not compact in $G_{j+1}$.  We know that
$T_i\setminus K_j$ will not be in $H$. Hence, all but one connected component
(the one that contains $H$) in $G_{j+1} \setminus T_i$ must belong to $\CT$.
Hence $T_i$ is not maximal. Contradiction.
\end{proof}

Let $\Gamma(\cdot)$ and $\Gamma^f(\cdot)$ denote the node neighbourhoods in the
faultless graph and the faulty graph respectively. It is easy to see the
following inequalities: $|\Gamma(T_i)| \geq { \alpha_e |T_i| \over \delta }$,
and $|\Gamma^f(T_i)| \leq \alpha_e \epsilon |T_i|$. These two inequalities
imply that $|\Gamma^f(T_i)| \leq \epsilon \delta |\Gamma(T_i)|$. Note that any
set $T_i$ was culled by {\em prune2} because its edge neighbourhood fell by a
factor of more than $\epsilon$.

The probability that the neighbourhood of some connected set $T_i$ in the
faulty graph went down from $\Gamma(T_i)$ to $\Gamma^f(T_i)$ is
(for the sake of brevity, $\Delta:=|\Gamma(T_i)|-|\Gamma^f(T_i)|$):
\begin{equation}
\label{eqn:prob}
\begin{split}
{ |\Gamma(T_i)| \choose |\Gamma^f(T_i)| } \cdot
p^{\Delta}
& \leq \left({ ep |\Gamma(T_i)| \over
\Delta }\right) ^\Delta\\
& \leq \left( { ep \over 1-\epsilon \delta } \right) ^{(1-\epsilon
\delta)|\Gamma(T_i)|}
\end{split}
\end{equation}
Note that this is valid under the condition that $ep+\epsilon \delta < 1$.  It
turns out that we have flexibility in bounding these two terms. We want to set
$\epsilon \delta$ closest to 1 so that degradation in expansion is minimal.
Therefore, if the following inequalities hold:
\[
\epsilon \delta \leq {1 \over 2 },\, ep \leq {1 \over 2 \delta^{4\sigma}},
\]
then the probability that $T_i$ is culled by {\em prune2} is at most
$\delta^{-3 \sigma |\Gamma(T_i)|}$ (this is an upperbound on the RHS in
\ref{eqn:prob}).
\[
\mbox{Pr} [ T_i \mbox{ is culled}] \leq \delta^{-3 \sigma |\Gamma(T_i)|}
\]
We enumerate two cases on the size of the neighbourhood of $T_i$s. In case 1 we
argue that a $T_i$ with a large neighbourhood is unlikely with high
probability. In case 2 we show that if all $T_i$s have small neighbourhoods
then it is unlikely that $\Sigma_i |T_i|$ is more than $n\over 2$ with high
probability. So, in case 2 assume that $|\bigcup_{i=1}^l T_i| \geq n/2$.  Let
$k=3 \log_\delta n$ in the following cases:

\noindent{\em Case 1:} $\exists i,\, |\Gamma(T_i)| \geq k$.\\
We know from before that the probability that a given compact subgraph $T_i$ is
culled is at most $\delta^{-3 \sigma |\Gamma(T_i)|}$.  We multiply this
probability with the number of ways of choosing such a subgraph. This gives us
the probability that there is a $T_i$ with such a large neighbourhood.  Each
compact subgraph has its corresponding perimeter. Therefore, the number of
compact subgraphs with boundary $|\Gamma(T_i)|$ is at most the number of
$\sigma \cdot |\Gamma(T_i)|$ sized spanning trees in the graph. This is at most
$n \cdot \delta^{2 \sigma \cdot |\Gamma(T_i)|}$. Note that
by definition, $\sigma \ge 1$. Hence,
\[
\begin{split}
\mbox{Pr} [ \exists T_i, |\Gamma(T_i)| > k] & \leq
\sum_{t=k}^{n} n \cdot \delta^{2 \sigma \cdot t} \cdot
\delta^{- 3 \sigma \cdot t}\\
& \leq n^2 \cdot \delta^{-k} \leq {1 \over n}
\end{split}
\]
\noindent{\em Case 2:} $\forall i,\, |\Gamma(T_i)| < k$.\\
\[
\mbox{Pr}[T_i\mbox{ is culled}] \leq \delta^{- 3\sigma \cdot
|\Gamma(T_i)|} \leq \delta^{-3}
\]
$T_i$s are disjoint by definition. Some $T_i$ and $T_j$ might share a bad node
in their neighbourhood leading to a dependency between them. But we do know
that since the perimeter of each $T_i$ is at most $k-1$, the maximum degree of
the dependency graph between the $T_i$s is $\delta \cdot (k-1)$.  Hence the
dependency graph can be coloured with $\delta \cdot (k-1) + 1 \leq \delta \cdot
k$ colours.

We know that $|\bigcup_{i=1}^l T_i| \geq n/2$. Hence there has to be a colour
class in the colouring of the dependency graph, let us call it $\CC$, such that
the $T_i$s in that colour class contain at least $n \over {2 \cdot \delta \cdot
k }$ nodes.

$|T_i| \leq { k \cdot \delta \over \alpha_e }$. Hence, the number of distinct
$T_i$s in $\CC$ has to be at least ${n \cdot \alpha_e \over {2 \cdot \delta^2
\cdot k^2 }}$. We know that the $T_i$s in $\CC$ are independent of each other.
We set a bound on $\alpha_e$ such that this probability becomes small.
Let $\alpha_e \geq {2 \delta^2 \cdot k^3 \over 9 n}$.
\[
\begin{split}
\mbox{Pr}[\forall T_i: T_i \mbox{is bad}] & \leq
\mbox{Pr}[\forall T_i\in \CC: T_i \mbox{ is bad}] \\
& \leq \delta^{-\frac{3 \alpha_e \cdot n}{2 \cdot \delta \cdot k^2 } } \leq
\delta^{-k/3} \leq {1 \over n}
\end{split}
\]
\[
\mbox{Pr}[ \mbox{nodes pruned} \geq n/2 ] \leq \mbox{Pr}[
\mbox{Case 1} ]+\mbox{Pr}[ \mbox{Case 2} ] \leq {2 \over n}
\]
\qed
\end{proof}

\subsection{Span of the mesh}


\begin{theorem}
\label{thm:d-mesh}
The $d$-dimensional mesh has span $2$.
\end{theorem}
\begin{proof}
Consider a compact set $S$ in the $d$-dimensional mesh $M$.  Let $B$
be the boundary nodes $\Gamma(S)$.  We place virtual edges
between nodes in $B$.  Two distinct nodes $u = (u_0, \ldots u_{d-1})$
and $v = (v_0, \ldots v_{d-1})$ have a virtual edge between them if
$|v_i - u_i| = 0$ for at least $d-2$ of its dimensions and $|v_i -
u_i| \leq 1$ for the rest.  Call the set of such virtual edges $E_v$.
In Lemma \ref{lm:1-comp}, stated below, we claim that the graph $(B,
E_v)$ is connected. Therefore, we can find a spanning tree for $B$
which has exactly $|B| - 1$ virtual edges. Since each edge in $E_v$
can be simulated by exactly 2 edges of $M$, we can say that there is a
spanning tree in $M$ for the nodes of $B$ with at most $2\cdot (|B| -
1)$ edges.
\end{proof}

\begin{lemma} \label{lm:1-comp}
Let $S \subset Z^d$ be a finite compact set, let $B$ be the boundary
nodes $\Gamma(S)$, and let $E_v$ be the set of virtual edges.  Then
the graph $(B,E_v)$ is connected.
\end{lemma}

\begin{proof}
We will show that for any two points $u$ and $v$ in $B$, there is a
path in $E_v$ connecting the two; if this can be done for every two
points, then $B$ is connected as we hope to prove.

Our proof uses some basic and standard homology theory of cell
complexes, which can be found in any introductory topology text; for
instance, see \cite{h-cit-79}.  Specifically, we use the $Z_2$ homology
of $d$-dimensional Euclidean space $R^d$.  We partition $R^d$ into a
complex of unit hypercube cells having the points of $Z^d$ as their
vertices.  Each $d$-dimensional unit hypercube cell has as its
boundary a set of $2d$ $(d-1)$-dimensional unit hypercube facets,
again having $Z^d$ as vertices, and so on.  In this complex, a {\em
$k$-chain} is defined to be any finite set of $k$-dimensional unit
hypercubes having points of $Z^d$ as vertices.  The boundary of a
$k$-chain $C$ is the symmetric difference of the boundaries of its
hypercubes; that is, it is the set of $(k-1)$-dimensional hypercubes
that are on the boundary of an odd number of the $k$-dimensional
hypercubes in $C$.  A {\em $k$-cycle} is defined to be a $k$-chain
that has an empty boundary, and a {\em $k$-boundary} is defined to be
a $k$-chain that is the boundary of some $(k+1)$-chain.  For quite
general classes of cell complexes in more complicated topological
spaces than $R^d$, every $k$-boundary is a $k$-cycle, but in $R^d$,
the reverse is also known to be true: every $k$-cycle is a
$k$-boundary.

Now, given $u$ and $v$, since $S$ is connected we can find a path
$p_1$ connecting $u$ to $v$ by a sequence of adjacent points in $S$.
We also find an edge $e_1$ connecting $u$ to an adjacent point of
$Z^d$ outside $S$, an edge $e_2$ connecting $v$ to an adjacent point
of $Z^d$ outside $S$, and a path $p_2$ connecting these two exterior
points by a sequence of adjacent points outside $S$ (since the
complement of $S$ is connected).  The union of $p_1$, $p_2$, and
$\{e_1,e_2\}$ forms a 1-chain in the cubical complex described above.
Moreover, this is a 1-cycle, because it has degree two at every vertex
it touches.  Therefore, it is the boundary of a 2-chain $C$; that is,
$C$ is a set of squares and $p_1\cup p_2\cup\{e_1,e_2\}$ is the set of
edges in the cubical complex that touch odd numbers of squares in $C$.

Next, let $U$ be the subset of $R^d$ formed by a union of axis-aligned
unit hypercubes, one for each member of $S$, and having that member as
its centroid; note that these hypercubes do not have integer vertices.
Let $B$ be the boundary facets of $U$; $B$ consists of a collection of
$(d-1)$-dimensional unit hypercubes that again do not have integer
vertices.  Finally, let $G=B\cap C$.

Whenever a square $s$ of $C$ and a $(d-1)$-dimensional hypercube $h$
of $G$ meet, they do so in a line segment of length $1/2$, that
connects the centroid of $h$ (where it is crossed by one edge of the
square) to the centroid of one of its boundary $(d-2)$-dimensional
hypercubes.  Thus $G$, the union of these line segments, can be viewed
as a graph that connects vertices at these points.  The degree of a
vertex at the centroid of $h$ is equal to the number of squares of $C$
that touch that point, and the degree of the other vertices can only
be two or four depending on which of the four vertices of the square
defining the vertex is interior to $U$.

Since the boundary of $C$ crosses $B$ only on the two edges $e_1$ and $e_2$,
these two crossing points have odd degree and all the other vertices of $G$
have even degree.  Any connected component of any graph must have an even
number of odd-degree vertices, so the two odd vertices $e_1\cap B$ and
$e_2\cap B$ must belong to the same component and can be connected by a
path $p_3$ in $G$.

Each length-$1/2$ segment of $p_3$ belongs to the boundary of a single
hypercube in $U$,
which has as its centroid a point of $B$.  Let $p_4$ be the sequence of
centroids corresponding to the sequence of edges in $p_3$.  Then $p_4$
starts at $u$, and ends at $v$.  Further, at each step from one edge in
$p_4$, either the current point in $B$ does not change, or it changes from
one point in $B$ to an adjacent point (when the corresponding pair of edges
in $p_4$ form a $180^\circ$ angle on two adjoining hypercubes), or it
changes from one point in $B$ to a point at distance $\sqrt{2}$ away
(when the corresponding edges in $p_4$ form a $270^\circ$ angle across a
concavity on the boundary of $U$).

So, we have constructed a path in $E_v$ between an arbitrarily chosen pair
of points $u,v$ in $B$, and therefore the graph $(B,E_v)$ is connected.
\end{proof}

Theorem~\ref{thm:d-mesh} implies that the $d$-dimensional mesh can
sustain a fault probability inversely polynomial in $d$ and still have
a large component whose expansion is no more than a factor of $d$
worse than the original.


\section{Conclusion}
\label{sec_conclusion}

In this paper we presented a general technique for determining the
robustness of the expansion of different networks both for adversarial
and random faults.  For random faults we have come up with a new
parameter, the span, which allows us to prove a strong result
regarding the robustness of high dimensional meshes. Among other
things, this result can provide useful insights into the working of
peer-to-peer networks like CAN \cite{RFH+01} which behaves like a
$d$-dimensional mesh in its steady state. Basically we have shown that
CAN can tolerate a fault probability which is inversely polynomial in
its dimension without losing too much in its expansion properties.

For the 2-dimensional mesh our result is related to the line of
research followed by Raghavan~\cite{Rag89}, Kaklamanis
et. al.~\cite{KKLMRRT90} and Mathies~\cite{Mat92} who show that
despite a constant fault probability (of as high as 0.4) a mesh with
random failures can emulate a fault free mesh using paths with stretch
factor at most $O(\log n)$. Since the distance of nodes in a graph of
expansion $\alpha$ is $O(\alpha^{-1}\log n)$~\cite{LR88}, our
technique gives essentially the same result albeit with a lower fault
probability. Additionally for meshes of constant dimension greater
than 2 our results imply a $O(\log n)$ dilation for path lengths, and
hence a way to generalize these earlier results to higher dimensions.

The strength of our technique is that it is able to yield results for
the 2-dimensional mesh which are comparable to previous results while
giving new results for higher dimensional meshes and providing a
general method suitable for analyzing any network whose span can be
estimated.

\subsection*{Open problems}

We conjecture that the butterfly, shuffle-exchange, and deBruijn
network all have a span of $O(1)$, which means that they can tolerate
a constant fault probability. Though the span may provide tight
results for these networks, the exponential dependency of the fault
probability on the span does not really give useful results if the
span is beyond $\log n$. Hence, either a better dependency result is
needed or a parameter better than the span is needed.  Clearly, as
mentioned in the introduction, having a parameter that can accurately
describe the fault tolerance of graphs w.r.t. expansion under random
faults would be very useful for many applications.




\bibliographystyle{abbrv}
\bibliography{faults}

\end{document}